\documentclass[a4paper,11pt]{article}
\pdfoutput=1
\usepackage{jheppub}
\usepackage{amsmath,amssymb}
\usepackage{bbold}
\usepackage{graphicx}
\usepackage{float}
\usepackage{color}
\usepackage [utf8]{inputenc}
\usepackage{subfigure}
\usepackage{euscript}
\usepackage{placeins}
\usepackage{diagbox}
\usepackage[thinlines]{easytable}

\allowdisplaybreaks[2] % permit pagebreak in the gather or the align enviroment

\def\OO{\mathcal{O}}
\def\bb{{\boldsymbol b}}
\def\ba{{\boldsymbol a}}
\def\be{{\boldsymbol e}}
\def\bx{{\boldsymbol x}}

\def\alphas{\alpha_{\mathrm{s}}}

\def\gsqa{g_{3\mathrm{d}}^2a}
\def\gsqb{g_{3\mathrm{d}}^2 b_\perp}
\def\gsq{g_{3\mathrm{d}}^2}
\def\gfour{g_{3\mathrm{d}}^4}
\def\gsix{g_{3\mathrm{d}}^6}

\def\mD{m_{\mathrm{D}}}
\def\Tr{\mathrm{Tr}}
\def\bp{{\bb_\perp}}
\def\bbp{b_\perp}
\def\bps{b_\perp^2}
\def\bpc{b_\perp^3}
\def\bpf{b_\perp^4}

\def\Cbp{C(\bp)}
\def\Wlb{\tilde W(L,\bp)}
\def\xc{x_\mathrm{cont}}
\def\yc{y_\mathrm{cont}}
\def\tension{\sigma_{\mathrm{EQCD}}}
\def\Eq#1{Eq.~(\ref{#1})}

\title{Transverse momentum broadening from the lattice}
\author{Guy D.\ Moore, Niels Schlusser}
\affiliation{Institut f\"ur Kernphysik, Technische Universit\"at Darmstadt\\
Schlossgartenstra{\ss}e 2, D-64289 Darmstadt, Germany}
\emailAdd{guy.moore@physik.tu-darmstadt.de, \\ nschlusser@theorie.ikp.physik.tu-darmstadt.de}

\abstract{
We present a calculation of the transverse momentum broadening of a
high-energy fundamental-representation particle, as calculated
nonperturbatively within EQCD using the technique
proposed by Caron-Huot and pioneered by Panero, Sch\"afer, and Rummukainen.
Our results are continuum extrapolated and provided at four
temperatures: 250 MeV, 500 MeV, 1 GeV, and 100 GeV.
}

\keywords{quark-gluon plasma, dimensional reduction, effective
  theories, kinetic theory, lattice gauge theory}

\begin{document}
\maketitle

\section{Introduction}
\label{sec:intro}

Computing transport coefficients of the Quark-Gluon Plasma (QGP)
directly from Quantum Chromodynamics (QCD) remains a major challenge
in theoretical particle physics. This is mainly due to transport being
an intrinsically real-time phenomenon while most of the methods rely
on the Euclidean time formalism. Transport coefficients of interest
are for instance the shear viscosity over entropy density ratio
$\frac{\eta}{s}$ \cite{Meyer:2007ic} or the jet broadening coefficient
$\hat{q}$ \cite{Bass:2008rv}. At high temperatures, one expects the
coupling to become small and QCD to become perturbative due to
asymptotic freedom \cite{Gross:1973,Politzer:1973}. However, it has
been found that this effect is compensated by highly occupied, and
therefore strongly interacting, infrared fields \cite{Linde:1980ts}.

A major effort was undertaken to compute transport coefficients of QCD
plasmas perturbatively at leading order
\cite{Arnold:2000dr,Arnold:2003zc} and, more recently, at
next-to-leading order
\cite{CaronHuot:2007gq,Ghiglieri:2013gia}. All computations of
next-to-leading order transport phenomena commonly rely on the
transverse collision kernel $C(q_\perp)$, which represents the rate
per unit time for a particle to undergo a scattering which changes the
transverse momentum by $q_\perp$.  The total rate of transverse
scattering, and the related momentum-broadening parameter, are
\begin{equation}
  \label{Cqdefined}
  \Gamma_{\mathrm{scatt}} = \int \frac{d^2 q_\perp}{(2\pi)^2}
  C(q_\perp) \,,
  \qquad
  \hat{q} \equiv \int \frac{d^2 q_\perp}{(2\pi)^2} q_\perp^2 \;
  C(q_\perp) \,.
\end{equation}
One can Fourier transform $C(q_\perp)$ into $\Cbp$, with
$\bp$ representing transverse distance (the impact parameter).
Casalderry-Solana and Teaney showed that $\Cbp$ is determined by
the falloff with length $L$ of the log-trace of a Wilson loop with two
edges of length $\pm \bp$ and two lightlike edges of spatial
length $L$ \cite{CasalderreySolana:2007qw}.
Remarkably, this clearly Minkowski definition still
allows a calculation via Euclidean methods; Caron-Huot showed that, at
next-to-leading order in the coupling, the desired Wilson loop is
equivalent to a modified Wilson loop \cite{CaronHuot:2008ni}
in a 3-dimensional Euclidean theory called Electrostatic Quantum
Chromodynamics, or EQCD.  EQCD was
first postulated by
Braaten and Nieto \cite{Braaten:1995cm}. In high-temperature QCD, as
described by EQCD, only the gluon Matsubara-0-mode remains dynamical,
all other degrees of freedom can be integrated out and contribute to
the effective field theory parameters. While the spatial components of
the gluon field persist in EQCD with a modified coupling
$g_{3\mathrm{d}}^2$, the temporal field component $A^0$ is no longer
protected by gauge invariance from acquiring a screening mass
$m_{3\mathrm{d}}^2$. Thus $A^0$ turns into a scalar in the adjoint
representation of SU(3) with a mass $\mD$ and a quartic self-coupling
$\lambda$. A perturbative matching between full QCD and
EQCD has been carried out for all parameters up to at least two-loop
order, for instance in \cite{Kajantie:1997tt,Laine:2005ai}.

Caron-Huot showed that, at next-to-leading order in the coupling and
up to IR safe corrections, EQCD also describes correlation functions
of spacelike or lightlike operators at sufficient distance, and
therefore the correlation functions relevant in the Wilson loop which
determines $\Cbp$ (at least for $\bbp \gg 1/2\pi T$).
Because EQCD captures the nonperturbative IR dynamics which spoil the
convergence of perturbative computations for
thermodynamical quantities such as the pressure \cite{Shuryak:1977ut,Kapusta:1979fh,Toimela:1983407,Arnold:1994eb,Zhai:1995ac,Kajantie:2002wa}
and correlation lengths \cite{Hart:2000ha,Hietanen:2008xb}, we expect this EQCD approach to
work for $\Cbp$ wherever dimensional reduction
\cite{Appelquist:1981vg,Nadkarni:1982kb} works, which may be as low a
twice the QCD crossover temperature \cite{Laine:2003ay}.

We should exploit this opportunity to determine a key dynamical
property of QCD, highly relevant for jet quenching
\cite{Bass:2008rv,Baier:2008js} and transport coefficients \cite{Ghiglieri:2013gia},
by performing a nonperturbative determination of $\Cbp$ within
EQCD, on the lattice.  This would be tremendously assisted by a
complete determination of all $\mathcal{O}(a)$ corrections between
lattice and continuum EQCD, both for the Lagrangian and for the Wilson
loop operator in question.  The
renormalization up to linear order in the lattice spacing $a$ is
known analytically for almost all parameters of EQCD
\cite{Moore:1997np}. More recently, the renormalization of the
modified, so-called `Null Wilson lines' was determined
\cite{DOnofrio:2014mld}. The only missing
$\mathcal{O}(a)$-contribution to renormalization stemmed from the mass
$m^2_{3\mathrm{d}}$; we recently determined it numerically in
\cite{Moore:2019lua}. This development allows for precision studies of
EQCD free from any $\OO(a)$ discretization errors.

In this paper we undertake a comprehensive study of $\Cbp$ in
EQCD on the lattice.  We hope the resulting \textsl{nonperturbative}
information about transverse momentum diffusion can be helpful in the
future, to study jet energy loss and thermalization in the Quark-Gluon
Plasma at temperatures where dimensional reduction is useful.
The next section will define the problem more completely.
Section \ref{sec:lattice} will describe in detail our lattice
procedure.
Section \ref{sec:results} will present our results, and finally we
will end with a short discussion.  Tabulated results and covariance
matrices appear in an appendix.

\section{Formulation of the problem}
\label{sec:formulation}

Electrostatic QCD is a dimensionally reduced effective theory for
high-temperature QCD. At high temperatures $T$, the Euclidean path
integral extends in all spatial directions but is periodic in
Euclidean time with extent $\beta = 1/T$, and periodic/antiperiodic
boundary conditions for integer/odd-half-integer spin fields.  The
long-distance physics is dominated by zero-Matsubara modes of the
gauge fields, or equivalently, the most important IR physics comes
from $A^\mu$ fields which are constant across the temporal direction.
Fluctuations which vary in the temporal direction, including all
fermionic fluctuations, can be
integrated out and contribute via the EFT parameters.  The temporal
gauge field component $A^0$ can be interpreted as a scalar $\Phi$,
living in the adjoint representation of SU(3).
Gauge invariance no longer protects it from receiving a mass; both a
mass and a quartic interaction are generated when we integrate out
fluctuations, leading to a continuum action which reads
\begin{equation} 
\label{cont_action}
S_{\mathrm{EQCD,c}} = \int \mathrm{d}^3x \, \left( \frac{1}{2 \gsq} \Tr \,
 F^{ij} F^{ij} + \Tr \, D^i \Phi D^i \Phi + \mD^2 \Tr \, \Phi^2 +
  \lambda \big( \Tr \, \Phi^2 \big)^2  \right) \, .
\end{equation}

\begin{figure}[htbp!] 
\centering
\includegraphics[width=\textwidth,keepaspectratio]{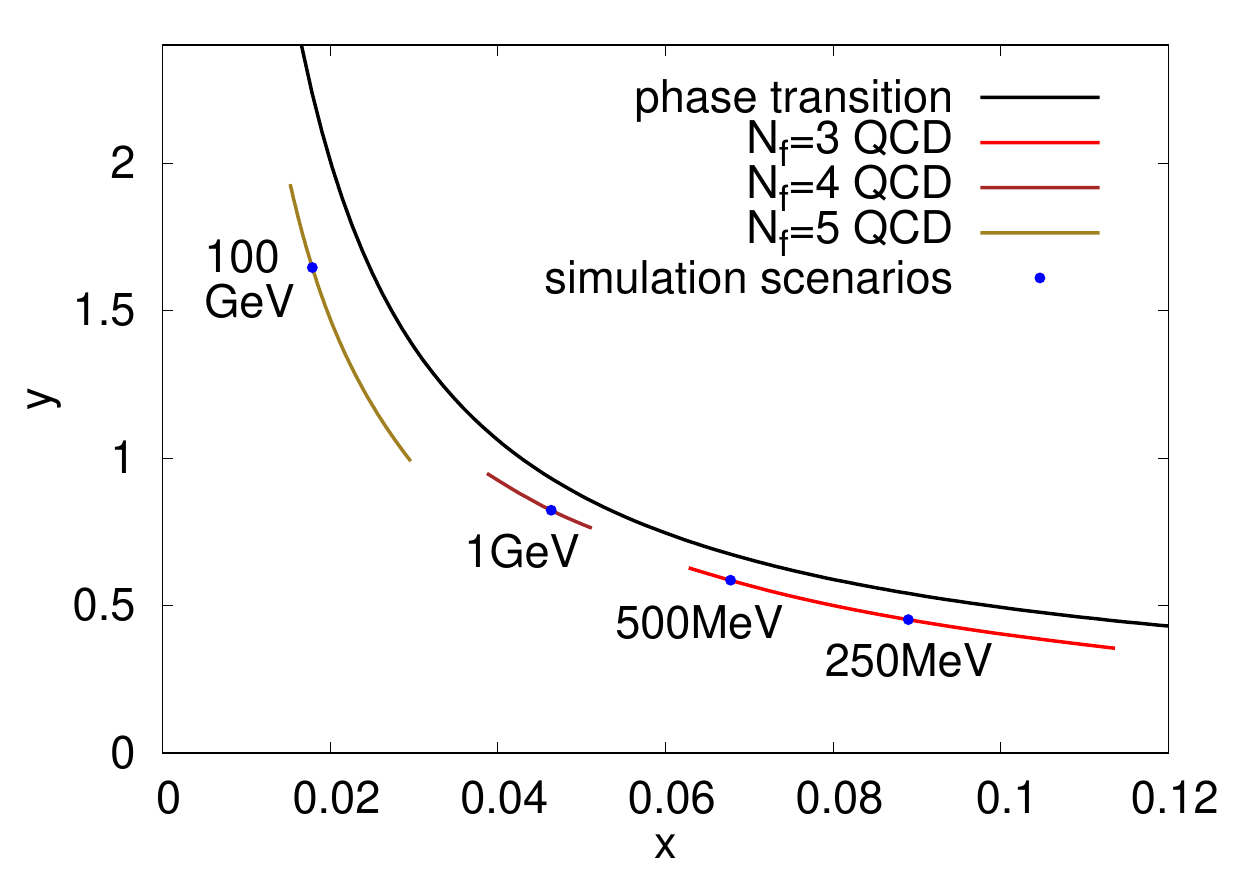}
\caption{$xy$ phase diagram of EQCD with the lines on which
  EQCD matches full QCD at Temperature $T$ and $N_\mathrm{f}$
  quark flavors. The phase transition if of first order in the
  physically interesting region \cite{Kajantie:1998yc}, and the
  physically relevant points lie in the supercooled ``high-temperature''
  symmetric phase below the transition line. The blue points mark our
  scenarios of interest from Table \ref{match_scenarios}.}
\label{phases_EQCD} \label{fig:xyplane}
\end{figure}

The gauge coupling $\gsq$ is dimensionful, and approximately equals
$g_4^2 T$ with $g_4^2$ the squared gauge coupling of the original 4D
theory.  Therefore one can consider $\gsq$ to set a scale (an energy
or inverse length scale), and we can use it to express the other
couplings in terms of dimensionless ratios:
$x \equiv \lambda / \gsq$, which expresses the strength of the scalar
self-coupling, and
$y \equiv \mD^2(\bar\mu=\gsq) / \gfour$, which expresses
how heavy the scalar mass is.  Formally $x \propto \alphas$
and $y \propto \alphas^{-1}$, so when the coupling is perturbative we
are in the small $x$ and large $y$ regime.  The relation between the
parameters $x,y$ and the gauge coupling and number of light fermions
of full QCD have been worked out to the two-loop level
\cite{Laine:2005ai} and are illustrated in Figure \ref{fig:xyplane}.
In this paper we will consider four specific cases, listed in Table
\ref{match_scenarios}.

\begin{table}[htbp!] 	
\centering
\begin{tabular}{|c|c|c|c|}	
\hline
$T$ & $n_{\mathrm{f}}$ & $x$ & $y$ \\
\hline
$250 \, \mathrm{MeV}$ & 3 & $0.08896$ & $0.452423$ \\
$500 \, \mathrm{MeV}$ & 3 & $0.0677528$ & $0.586204$ \\
$1 \, \mathrm{GeV}$ & 4 & $0.0463597$ & $0.823449$ \\
$100 \, \mathrm{GeV}$ & 5 & $0.0178626$ & $1.64668$ \\
\hline
\end{tabular}
\caption{3D EQCD parameters for four typical scenarios.}
\label{match_scenarios}
\end{table}

Our observable of interest is the modified Wilson loop proposed in
\cite{CaronHuot:2008ni},
\begin{align}
  \label{Cdef}
  \Cbp &= -\lim_{L \to \infty} \frac{1}{L} \ln \Tr \,
  \tilde W_{(0,0);(L,0);(L,\bp);(0,\bp)} \\
  &\equiv -\lim_{L \to \infty} \frac{1}{L} \ln \Wlb  \,,
\end{align}
where $\tilde W$ denotes the modified (fundamental representation) Wilson
loop with the corners
$(0,0),\,(L,0),\,(L,\bp),\,(0,\bp)$, written in terms of the value
in the $z$ direction and the transverse direction.
By ``modified'' Wilson loop, we mean that the longitudinal components
contain both the gauge field and the exponential of the $\Phi$ field:
\begin{eqnarray}
\hspace{-2em}  \Wlb &=& \text{P exp} \left(
  \int_0^L \Big[-iA_z(z,0) + \Phi(z,0) \Big] dz
  + \int_0^\bp (-iA_\perp(L,\bp')) d\bp'
  \right. \nonumber \\ && \phantom{\mathrm{Pexp} ()} \left.
  + \int_L^0 \Big[ +iA_z(z,\bp) - \Phi(z,\bp) \big] dz
  + \int_\bp^0 (+iA_\perp(0,\bp')) d\bp' \right) \,.
\end{eqnarray}
Note that $\Phi$ appears without a factor $i$, that is, the long edges
of the Wilson loop are not unitary.  This term can be understood as
$-iA_0$ in the original Minkowski picture, with $A_0$ rewritten as
$\Phi$ and rotated by a factor $i$ when we pass to Euclidean
signature.  Our goal is to establish the value of $\Cbp$ using
\Eq{Cdef} for several transverse distances, in each of the four cases
listed in Table \ref{match_scenarios}.

\section{Computational details}
\label{sec:lattice}

\subsection{Lattice implementation}

The EQCD continuum action was discretized on a three-dimensional spatial lattice,
including the now-complete $\OO(a)$ corrections to the lattice
parameters.  The lattice fields are updated using a mixture of
heatbath and overrelaxation updates.  One full update consists of a
sweep over all sites of one heatbath update to each scalar and link,
followed by four sweeps in which each scalar and link is updated with
over-relaxation. The presence of an adjoint scalar
requires an additional accept-reject step be added to the gauge boson
updates.  Our code involves custom modifications of the openQCD-1.6
codebase \cite{openQCD}.  For more details of our lattice
implementation, we refer to \cite{Moore:2019lua}.

The modified Wilson loop \eqref{Cdef} on the lattice is a local
observable which suffers from significantly higher noise levels than
volume-averaged observables such as the scalar condensate $\frac{1}{V}
\sum_\bx \Tr \, \Phi^2(\bx)$. This effect is especially strong for
loops that enclose large areas, which automatically applies in our
case since we would like to extract an observable that requires an
extrapolation of $L \to \infty$. An algorithm that was designed to
overcome that problem is the multilevel algorithm proposed by
L\"uscher and Weisz in 2001 \cite{Luscher:2001up,Meyer:2002cd}. It
relies on freezing one or multiple surfaces perpendicular to the
Wilson loop and updating the subvolumes separately. This allows to
average over the subvolumes independently and measuring the final
observable as a correlation of multiple, pre-averaged quantities,
which reduces the noise drastically. Originally, this algorithm was
designed for pure SU(3) Yang-Mills theory, but it found an
application to EQCD, too \cite{Panero:2013pla}. We split our lattices
along the largest extend $N_z$ in 4 sublattices, on which we perform
80 update sweeps separately before an update sweep through the
complete volume is conducted.

Following Panero, Sch\"afer and Rummukainen \cite{Panero:2013pla}, the
lattice implementation of the modified Wilson loop reads
\begin{align}	\label{latt_Wdef}
\Wlb &= \Tr \, \left( \tilde{U}_{(0,0);(L,0)} U_{(L,0);(L,\bp)}
\tilde{U}^{-1}_{(0,\bp);(L,\bp)} U^\dagger_{(0,0);(0,\bp)} \right) \\
\tilde{U}_{x;x+(L,0)} &= \prod^{n_\mathrm{L}-1}_{n=0} U_3 \left( x +
an \be_z \right) \exp \left( - Z \, \Phi (x + a(n+1) \be_z) \right)
\notag \\
U_{x;x+(0,\ba)} &= \prod_{n=0}^{n_\ba - 1} U_1 \left( x + an \be_x
\right) \notag \, ,
\end{align}
where $U_i$ is the standard gauge link in $i$-direction, the transverse
separation is assumed to be $n_\bb$ lattice spacings $a$ in the
x-direction $\bb = n_\bb a \be_x$ and $Z$ is the renormalization
factor of the Null Wilson line.
This quantity was analytically computed to $\OO(a)$ in
\cite{DOnofrio:2014mld}. We repeat their main result for SU(3)
explicitly for convenience:
\begin{equation}
  \frac{Z^2}{Z_\Phi \gsqa} = 1
  + \gsqa \left( \frac{\Sigma}{4 \pi} - \frac{3 \xi}{\pi} \right)  \, ,
\end{equation}
where $\gsqa$ is the dimensionless lattice spacing, $Z_\Phi$ is the
overall normalization factor of the scalar $\Phi$, which we set
$Z_\Phi=1$ without loss of generality, and $\Sigma=3.17591153562522$
and $\xi=0.152859324966$ are standard integrals in the
lattice-continuum-matching of EQCD.

When computing a Wilson loop it is often possible to replace each link
with its heat-bath average in computing the Wilson loop's value.  This
is possible so long as no term in the lattice action contains more
than one link which appears in the Wilson loop.  Therefore we cannot
apply this method here, because the scalar field appears in an action
term containing the link variable and it also appears in the modified
Wilson loop.  Similarly, we have not found an efficient way to average
over scalar fields, because each scalar field depends on the
neighboring scalar fields and because of the quartic term in the
action.  Therefore the multilevel procedure is the only noise
reduction technique we have applied.

\subsection{Extracting $\Cbp$}

The previous section explains how we obtain data for $\Wlb$ at
multiple lengths $L$, transverse distances $\bb$, and lattice spacings
$\gsqa$, each at four ``temperatures'' ($x,y$ choices).  Here we
explain how we use this data to extract $\Cbp$.
There are three different limits that have to be taken into account properly:
\begin{enumerate}
\item infinite volume limit $V \to \infty$
\item infinite length limit $L \to \infty$
\item continuum limit $a \to 0$
\end{enumerate} 
Since EQCD possesses a mass gap, the first limit can be easily reached
by choosing sufficiently large volumes \cite{Hietanen:2008tv}. The
continuum limit is performed by a standard polynomial extrapolation of
$\gsqa \to 0$, where the linear term was eliminated by the full
$\OO(a)$ improvement. So the remaining limit to be treated is the
infinite length limit, which we will discuss in the next few
paragraphs.

Unfortunately, the information about $\Cbp$ in $\Wlb$ is contaminated
by (in principle infinitely many) higher states' energies
\cite{Gattringer:2010zz}
\begin{equation}
\Wlb = c_0 e^{- L \, \Cbp} + \sum_{n=1}^\infty c_n e^{- L \, E_n(\bp)} \, ,
\end{equation}
where $c_i$ are prefactors resulting from the geometry of the Wilson
loop and lattice artifacts and are not important for our
purpose. Fitting a large number of exponential functions is in general
a very hard problem, which gets even worse if the decay constants are
numerically close to each other. What saves the day is that the
energies are increasingly ordered in $n$, ie.\ their exponential decay
happens faster at large $L$ as $n$ increases and we are only
interested in the lowest energy $\Cbp$. Conversely, the relative error
of a Wilson loop $\Wlb$ scales inversely with the enclosed area, so
small loops feature good statistics. It is crucial to find a regime in
which the balance between sufficiently small Monte Carlo errors and
sufficiently large $L$ for small contamination is maintained.
In our case, $\gsq L \geq 1.0$ fulfilled that requirement such that it
is sufficient to consider only one higher state. Thus, the fit
function for the $L \to \infty$ extrapolation reads
\begin{equation}
\Wlb = c_0 e^{- L \, \Cbp} + c_1 e^{- L \, E_1(\bp)} \, .
\end{equation}

\begin{figure}[htbp!] 
\centering
\includegraphics[width=\textwidth,keepaspectratio]{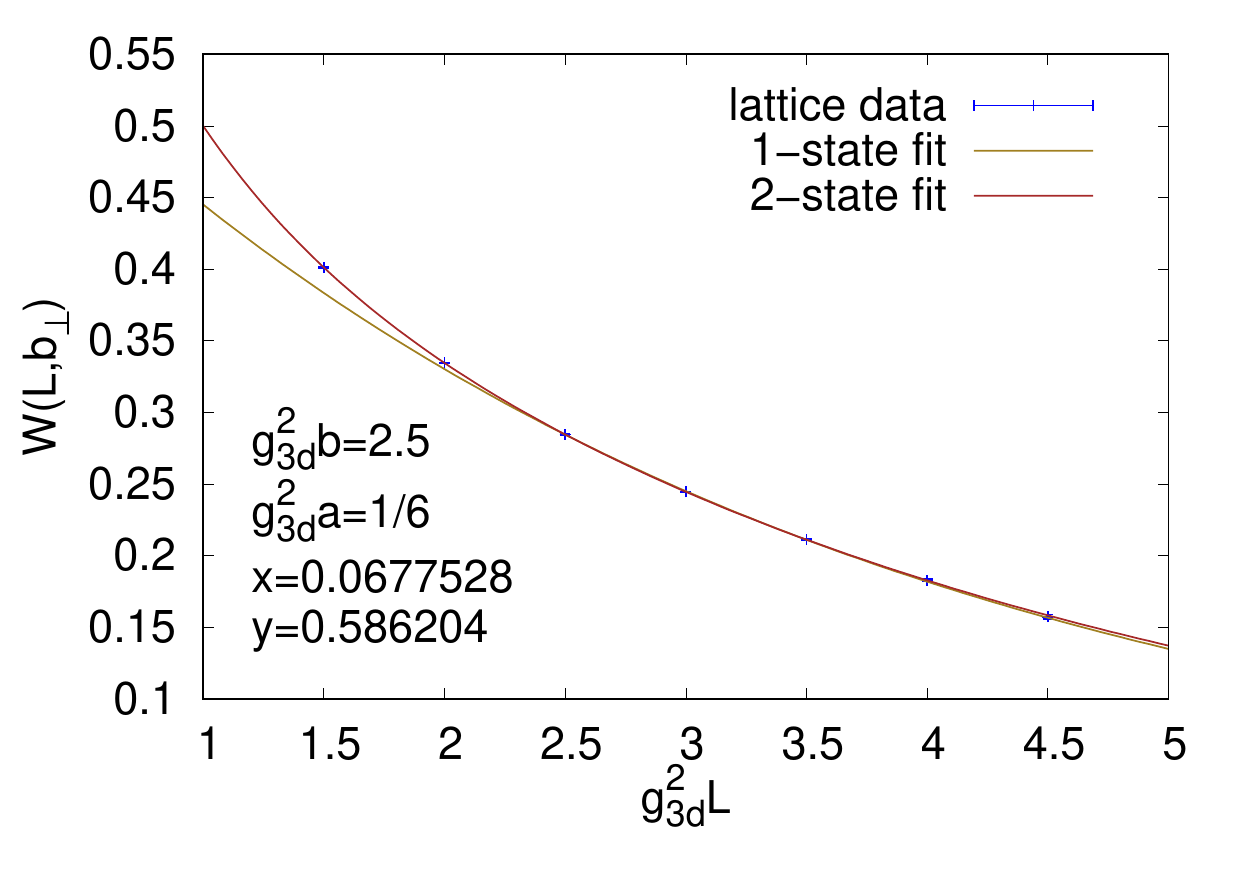}
\caption{Genuine $L \to \infty$ fit for $\gsqb=2.5$ at
  $\gsqa=1/6$ and $x=0.0677528$. Note that this figure only
  illustrates the initial guess finding. The actual values
  extracted from the data by the variable projection cannot be
  displayed, since this procedure does not give values for the
  $c_i$'s. However, we found that the final results from the
  variable projection were very close to the initial ones
  determined as in the plot.} 
\label{l_extra}
\end{figure}

Exponential fits of this form are in general very unstable. There are
a few techniques one can apply to improve that, however. Firstly, one
can estimate starting values close to the final fit values, for which
a procedure was outlined in \cite{Bazavov:2019www}. The second method
we apply is the so-called variable projection
\cite{OLeary2013}. Roughly speaking, one gives up on determining the
$c_i$'s and finds the minimum of $\chi^2$ in the reduced parameter space, only. In
our case, this is an appropriate procedure since we are not interested
in the values of the $c_i$'s, anyway.

As a last obstacle, we determine our data for all $\Wlb$ at a given
lattice spacing from the same ensemble, which means that our data is
highly correlated, not only along the Monte Carlo time axis, but also
for the different $L$ and $\bp$. The correlation along the Monte Carlo
time axis can be eliminated by binning the data; the bin size has to be
varied until a plateau for the errors is reached. The correlation of
different lengths is taken care of by performing correlated,
variable-projected fits \cite{Beane:2010em,OLeary2013}. Last but not
least, the correlation for the final, continuum-extrapolated different
$\bp$ is less severe than the one for the different lengths since
multiple lattice spacings (ensembles) contribute to the
continuum-extrapolated points of $\Cbp$. Nevertheless, we report the
covariance matrices for all $\Cbp$ at our four temperatures in
App.~\ref{app1}.

\begin{figure}[htbp!] 
\centering
\includegraphics[width=\textwidth,keepaspectratio]{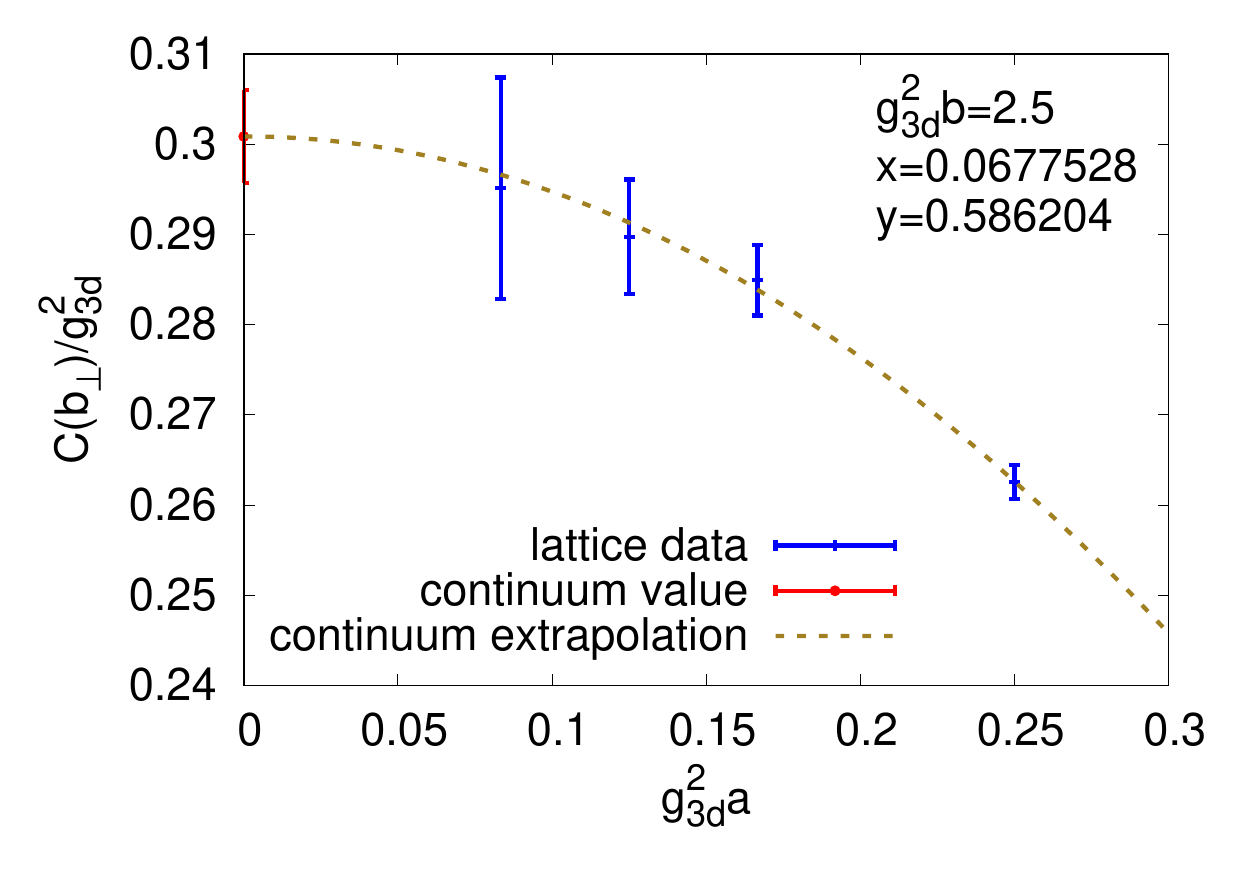}
\caption{Genuine continuum extrapolation for $\gsqb=2.5$ at $x=0.0677528$.}
\label{cont_extra}
\end{figure}

This data can now be extracted at various lattice spacings and
extrapolated to the continuum, cf.\ Fig.~\ref{cont_extra}. We choose the
lattice spacings such that a quadratic interpolation is
sufficient. Since the linear term is eliminated by the full $\OO(a)$
improvement, all continuum extrapolations are fits with only two free
parameters, improving the error of the extrapolated value
drastically. The shape of the extrapolation fit in
Fig.~\ref{cont_extra} is clearly quadratic, confirming that our
improvement procedure succeeded in eliminating all linear-in-$a$
renormalizations and rescalings.

\section{Results}
\label{sec:results}

\subsection{Analytical expectations: small $\bp$}
\label{analyticsmall}

Before presenting numerical results, we should start by asking, what
answers do we expect?  Of course we don't know what the behavior of
$\Cbp$ should be, otherwise there would be no need to measure it
nonperturbatively on the lattice.  But in limiting regimes, namely
$\gsqb \ll 1$ and $\gsqb \gg 1$, we might expect simpler
behavior.

Let us start with $\gsqb  \ll 1$.  First note that $\Cbp$ has the
same units as energy.  To see this note that the log trace of a Wilson
loop is dimensionless, so \Eq{Cdef} shows that $\Cbp$ has units of
inverse length or energy.  Alternately, $C(q_\perp)$ describes a
probability per unit length and momentum-squared, and so has units of
inverse energy.  Fourier transforming $\int d^2 q_\perp$ introduces
two factors of energy, making $\Cbp$ linear in energy.  Next, note
that both $\gsq$ and $\mD$ have units of energy.
Perturbation theory is an expansion in $\gsq$, but since this quantity
is dimensionful it should be balanced by powers of something else with
dimensions; hence we have an expansion in $\gsqb$ and/or in
$\gsq/\mD$.  Therefore we formally expect that
the small-$\bbp$ expansion, as an expansion in loop order, is of form
\begin{equation}
  \label{Cbseries}
  \begin{array}{llllll}
    \Cbp \sim & {} \gsq &{} + \gsq \bbp \mD &{} + \gsq \bps \mD^2
    &{} + \OO(\bpc)   & \quad \mbox{LO} \\
  & &{} + \gfour \bbp &{} + \gfour \bps \mD & {} + \OO(\bpc)
  & \quad \mbox{NLO} \\
  & && {} + \gsix \bps & {} + \OO(\bpc)
  & \quad \mbox{NNLO}\,. \end{array}
\end{equation}
If we were computing the standard Wilson loop, the first two
leading-order terms might arise; but for the structure we do consider,
there are cancellations at leading order between $A_z$ and $\Phi$
contributions, precisely because $\Phi$ couples in an antihermitian
way.  Since the $\Phi$ field is heavy, the cancellation is incomplete
and the $\gsq \bps \mD^2$ term is present, but the $\gsq$ and
$\gsqb \mD$ terms are absent.  In particular, the leading
short-distance contribution is \cite{Aurenche:2002wq}
\begin{equation}
  \label{CLO}
  \Cbp_{\mathrm{LO}} = \frac{\gsq}{6\pi}
  \left( 1 - \gamma_\mathrm{E} + \ln(2) - \ln(\bbp \mD) \vphantom{\Big|}
  \right) \mD^2 \bps + \OO(\bpf \ln(\bbp)) \,,
\end{equation}
which as expected scales as $\gsq \mD^2 \bps$.  Note that a quadratic
term in $\Cbp$ can be understood, using \Eq{Cqdefined}, as
$\Cbp \simeq \frac{\hat{q}}{4} \bps$ (see \cite{Ghiglieri:2013gia}
Appendix C).  Therefore the logarithmic term here represents a log UV
divergence in the Coulombic value of $\hat{q}$, which is well known.
Also note that the dominant momentum region giving rise to \Eq{CLO}
is $q_\perp \sim \mD$, since this is the momentum region where the
cancellation between $A_z$ and $\Phi$ first breaks down.  The momentum
region $q_\perp \sim \gsq$ gives rise to an $\OO(\gsix \bps)$
contribution, which therefore indicates the order where
nonperturbative physics will enter.

At NLO the full $\bp \mD$ dependence has been worked out in Ref.\
\cite{Ghiglieri:2013gia}, based on $q_\perp$-space results from
\cite{CaronHuot:2008ni}.  They find
\begin{equation}
  \label{CNLO}
  \Cbp_{\mathrm{NLO}} = \Cbp_{\mathrm{LO}} - \frac{\gfour \bbp}{8 \pi}
  + \frac{3 \pi^2 +10 - 4 \ln 2}{32 \pi^2}
  \gfour \mD \bps + \OO(\gfour \mD^2 \bpc) \,.
\end{equation} 
The linear term is the only term linear in $\bbp$ which will arise, and
is therefore a clean prediction of perturbation theory.  The
second term is formally suppressed relative to the LO expression by
a factor $\sim \gsq/\mD$, indicating that in this region, perturbation
theory is an expansion in $(\gsq/\mD) \sim y^{-1/2}$.
Similarly, the unknown NNLO contribution is of order
$\gsix \bps$.  Unfortunately a 2-loop calculation would not be
sufficient to determine this term, since as we already discussed, this
is the order where perturbation theory for this quantity breaks down
due to IR divergences; all higher loop orders also contribute at this
order, so the $\gsix \bps$ coefficient is nonperturbative.  This
unknown nonperturbative entry is suppressed, with respect to the
leading-order result, by a factor of $y^{-1}$.  Therefore, the
highest-temperature case we consider, with $y=1.65$, should show
reasonable convergence and the two known perturbative terms should be
relatively close to determining the true linear plus quadratic
behavior at small $\bbp$.  But for the other values we consider,
perturbative results for the $\bps$ coefficient will not be useful and
this coefficient (and therefore $\hat{q}$) has to be fitted.  However
the $\bbp$ and $\bbp^2 \ln(\bbp\mD)$ terms are clean predictions of
perturbation theory.

\subsection{Analytical expectations:  large $\bbp$}
\label{analyticbig}

The large $\bbp$ region has been discussed by \cite{Panero:2013pla},
who argue that $\gsqb \gg 1$ corresponds to the region where
Wilson loops display area-law behavior.  The $\Phi$ field correlator
essentially vanishes between opposite edges of the Wilson line and
does not contribute to the $\bbp$ dependence in this regime; therefore
we expect
\begin{equation}
  \label{area_law}
  \Cbp \simeq \tension \: \bbp \, ,
\end{equation}
where $\tension \propto \gfour$ is the EQCD string tension.
The EQCD string tension was predicted in \cite{Laine:2005ai},
continuing the continuum-extrapolated result of three-dimensional pure
gauge theory \cite{Teper:1998te}, ie.\ magnetostatic QCD (MQCD) to
EQCD 
\begin{equation}	\label{EQCD_string_tension}
\frac{\sqrt{\tension}}{\gsq} = \left[ 1 - \frac{1}{48} \frac{3}{\pi \sqrt{y}} - \frac{17}{4608} \left( \frac{3}{\pi \sqrt{y}} \right)^2 \right] \times 0.553(1) \, ,
\end{equation}
applying the 2-loop perturbative matching between EQCD and MQCD 
from \cite{Giovannangeli:2003ti}.
However, this calculation relies on a perturbative matching of
the MQCD coupling to its EQCD equivalent, which is again a formal
expansion in $1/\sqrt{y}$ which may not show good convergence.
%Instead, we can compute the string tension of EQCD from our
%simulations; by setting $Z=0$ in \eqref{latt_Wdef} we recover
%the standard Wilson loop.  The trace of a standard Wilson loop should
%depend on its transverse and longitudinal extents, if both are large,
%as
%\begin{equation}
%  - W(L,\bp) = A + B ( 2L + 2\bbp )
%  + \tension L \bbp \,,
%\end{equation}
%where $\tension$ is the area-law or linear-confining
%coefficient, $B$ is a coefficient associated with perimeter-law
%contributions, and $A$ is a constant associated with the corners.
%Both $A$ and $B$ are expected to suffer from UV divergences and are
%therefore lattice-spacing dependent, but $\tension$ should have a
%valid continuum value, which can be obtained as
%\begin{equation}
%  \label{extractsigma}
%  \tension = - \lim_{\gsqa \to 0, \bbp \to \infty, L \to \infty}
%  \frac{\partial}{\partial \bbp}
%  \frac{\partial}{\partial L} W(L,\bp) \,.
%\end{equation}
%Therefore it is straightforward to predict the large $\bbp$ behavior
%of $\Cbp$, which grows linearly with a coefficient which is
%nonperturbative but can be independently evaluated.

\subsection{Numerical results}

\begin{figure}[ptbh] 
  \centering 
  \begin{tabular}{cc}
    \includegraphics[scale=0.5]{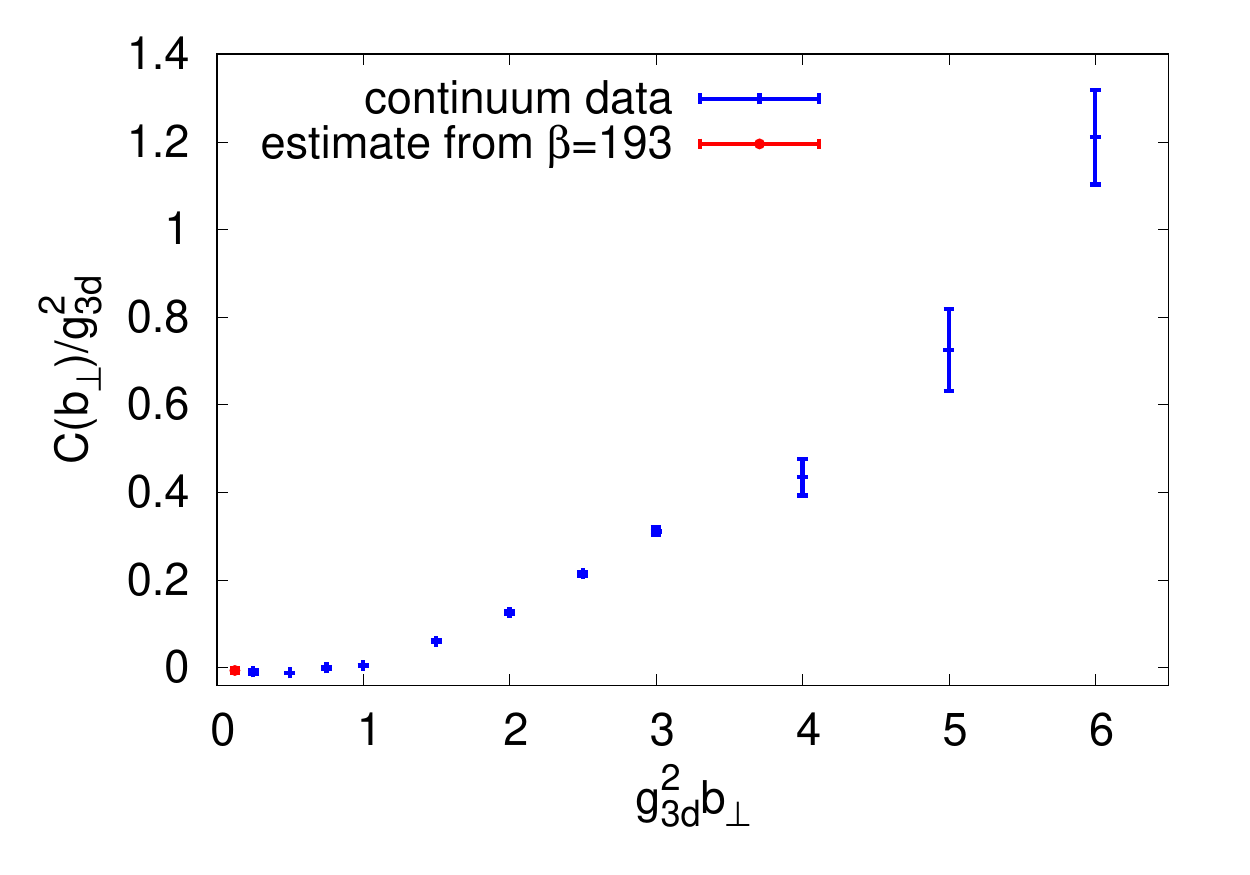} & 
    \includegraphics[scale=0.5]{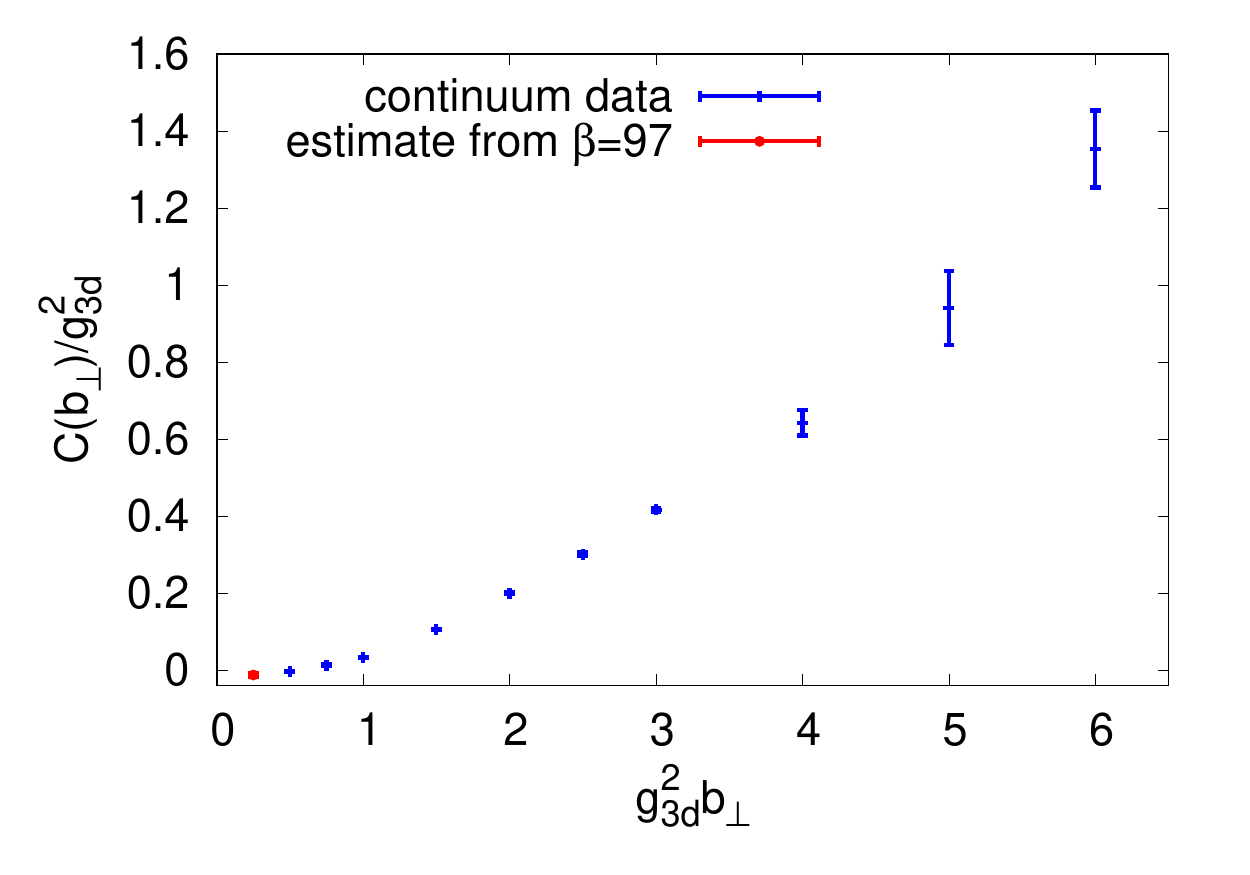} \\
    (a) $x=0.08896$ & (b) $x=0.0677528$ \\
    \includegraphics[scale=0.5]{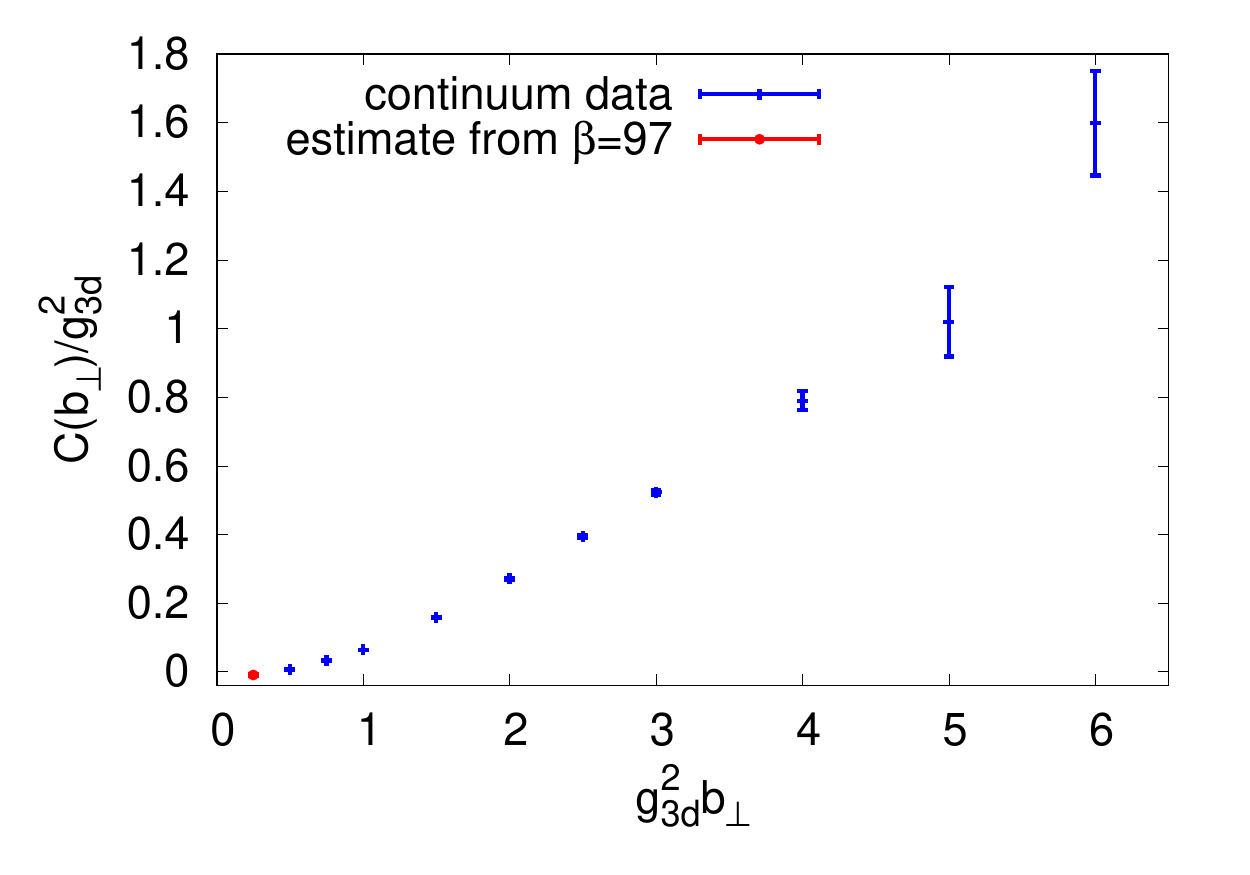} &
    \includegraphics[scale=0.5]{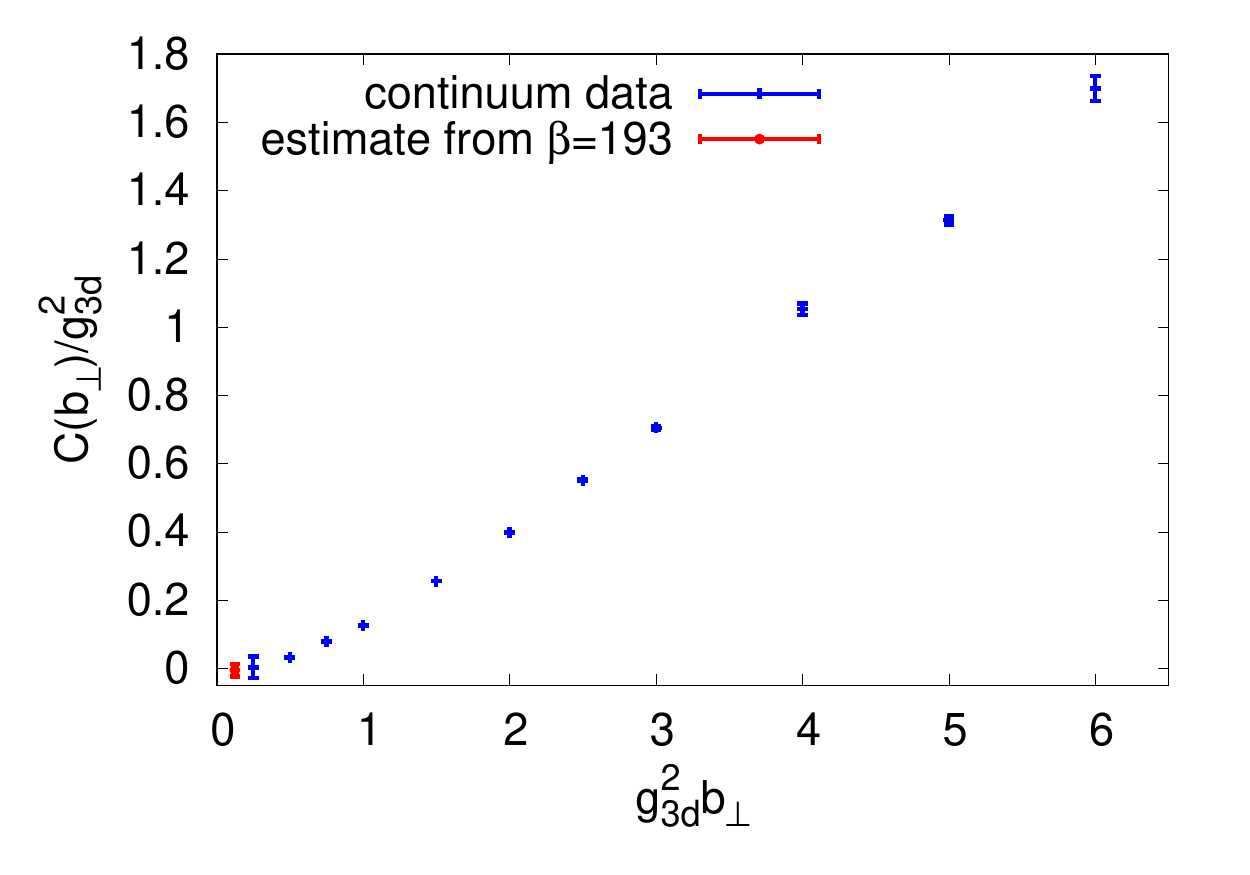} \\ 
    (c) $x=0.0463597$ & (d) $x=0.0178626$ \\
  \end{tabular}
  \caption{$\Cbp$ for different $x$.  The blue points are based on
    continuum extrapolations; the red points are from a single lattice
    spacing, as described in the text.}
	\label{cont_limits}
\end{figure}

Section \ref{sec:lattice} has already described how we extract
$W(L,\bp)$ values at each lattice spacing, and how we extrapolate
these to the large $L$ and small $\gsqa$ limits.
Doing so, we find the results tabulated in Table \ref{sim_res_tab}
and displayed in Fig.~\ref{cont_limits}. 
These represent our principal findings.

\begin{table}[ptbh] 
\centering {\small
\begin{tabular}{|c|c|c|c|c|}	
\hline
 & $\xc = 0.08896$ & $\xc = 0.0677528$ & $\xc = 0.0463597$ & $\xc = 0.0178626$ \\
 & $\yc = 0.452423$ & $\yc = 0.586204$ & $\yc = 0.823449$ & $\yc = 1.64668$ \\
 & & & & \\[-10pt]
$\gsqb$ & $\left. \frac{\Cbp }{ \gsq} \right\vert^{N_\mathrm f = 3}_{250~\mathrm{MeV}}$ & $\left. \frac{\Cbp }{ \gsq} \right\vert^{N_\mathrm f = 3}_{500~\mathrm{MeV}}$ & $\left. \frac{\Cbp }{ \gsq} \right\vert^{N_\mathrm f = 4}_{1~\mathrm{GeV}}$ & $\left. \frac{\Cbp }{ \gsq} \right\vert^{N_\mathrm f = 5}_{100~\mathrm{GeV}}$ \\
 & & & & \\[-10pt]
\hline
$0.125$ & $-0.0058(44)$ & - & - & $-0.005(19)$ \\
$0.25$ & $-0.0091(36)$ & $-0.0130(36)$ & $-0.0094(34)$ & $0.003(32)$ \\
$0.5$ & $-0.01154(63)$ & $-0.00394(87)$ & $0.00599(58)$ & $0.03166(49)$ \\
$0.75$ & $0.0000(11)$ & $0.0128(17)$ & $0.0337(10)$ & $0.07904(81)$ \\
$1.0$ & $0.00623(82)$ & $0.03313(61)$ & $0.06366(36)$ & $0.12649(33)$ \\
$1.5$ & $0.0606(17)$ & $0.1055(11)$ & $0.15803(94)$ & $0.25599(62)$ \\
$2.0$ & $0.1269(33)$ & $0.2002(28)$ & $0.2712(18)$ & $0.3986(12)$ \\
$2.5$ & $0.2150(41)$ & $0.3009(51)$ & $0.3947(32)$ & $0.5518(22)$ \\
$3.0$ & $0.3114(86)$ & $0.4164(46)$ & $0.5234(67)$ & $0.7048(45)$ \\
$4.0$ & $0.435(42)$ & $0.643(33)$ & $0.790(28)$ & $1.053(17)$ \\
$5.0$ & $0.726(94)$ & $0.941(96)$ & $1.02(10)$ & $1.314(13)$ \\
$6.0$ & $1.21(11)$ & $1.35(10)$ & $1.60(15)$ & $1.700(36)$ \\
\hline
$\hat{q} / \gsix$ & $0.1847(78)$ & $0.230(10)$ & $0.3637(60)$ & $0.6424(47)$ \\
\hline
$\sigma_\mathrm{EQCD} / \gfour$ & $0.2836(10)$ & $0.2867(10)$ & $0.2901(11)$ & $0.2952(11)$ \\
\hline
\end{tabular} }
\caption{
  Results for $\Cbp$ for four temperatures and a range of
  transverse separations.  All data points are continuum extrapolated,
  with errors representing all statistical and systematic errors
  associated with the data extraction and extrapolations, except for
  the first (smallest $\gsqb$) entry in each column; see text.
  We also quote the extracted value of $\hat{q}$ and the string
  tension as determined in \cite{Laine:2005ai}, see text.\protect\footnotemark
}
\label{sim_res_tab}
\end{table}

\footnotetext{In an earlier version of this paper, a numerical mistake 
  in our fitting procedure caused our results to be too small by a factor 
  of $1/2$.}

Note that every result in Table \ref{sim_res_tab} is based
on data from at least three lattice spacings, extrapolated to the
continuum -- \textsl{except} for the first (smallest-separation) data
point in each column, see Table \ref{sim_params} in the appendix.
This point is based on a single lattice
spacing.  Analyzing the lattice-spacing dependence of the points
where we can perform a continuum extrapolation, we find that the $a^2$
coefficient is fairly constant for small $\bbp$; so we assume that the
same $a^2$ extrapolation coefficient applies for this smallest-$\bbp$
point as for the next-smallest $\bbp$ data, and assign
100\% systematic errors to this estimate.  We then combine this
(systematic) error with the statistical error in quadrature.

\begin{figure}[phtb] 
  \centering
	\includegraphics[width=0.8\textwidth,keepaspectratio]{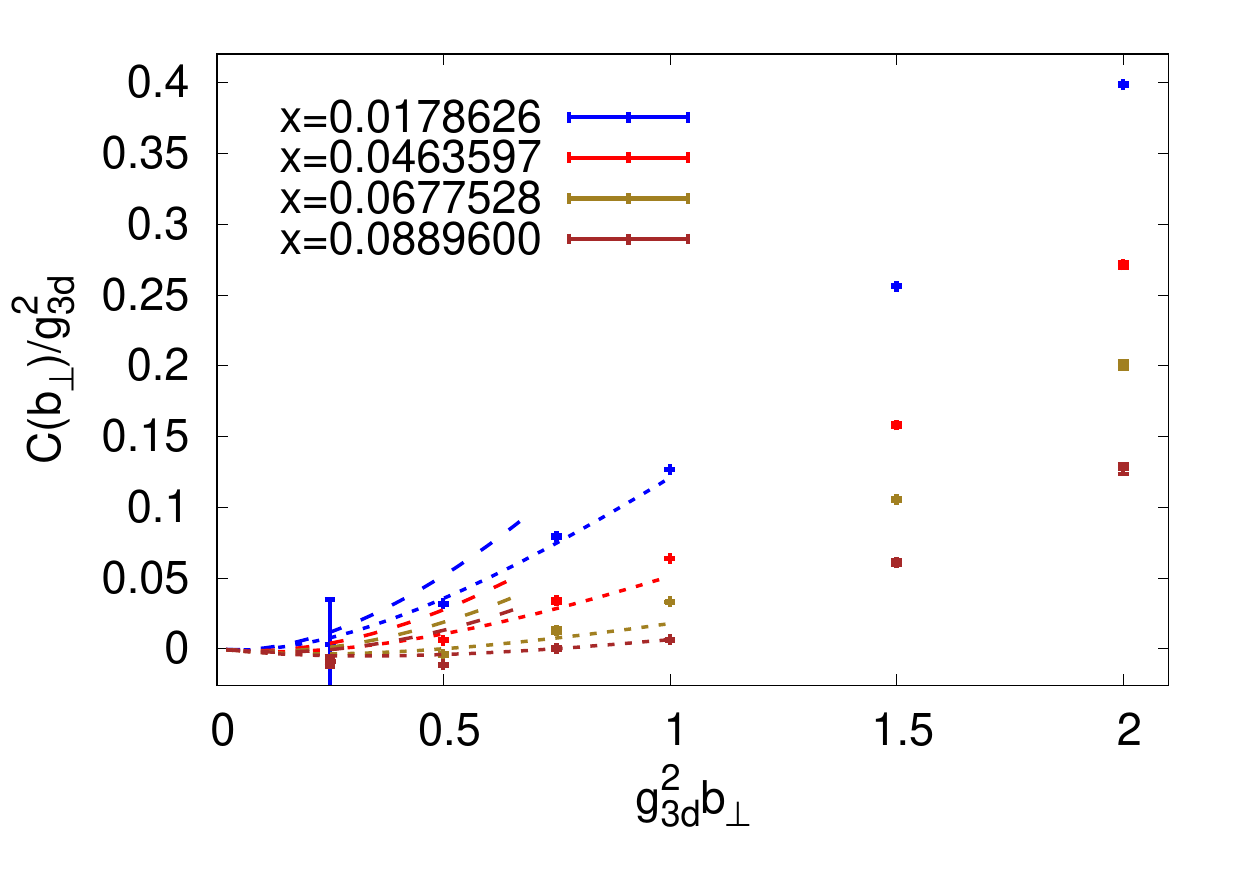} 	
  \caption{Comparison of $\Cbp$ for different $x$ at small
    $\gsqb$. The long dashed lines mark the perturbative results
    for $\Cbp$ at NLO, \Eq{CNLO}. The short
    dashed lines mark a quadratic fit to the datapoints satisfying
    $\gsqb \leq 0.75$, see text.}
	\label{small_gsqb}
\end{figure}

Let us examine the small $\bbp$ region in more detail.  As we saw in
Subsection \ref{analyticsmall}, we expect that $\Cbp$ scales, for
small $\bbp$, as
\begin{equation}
\label{qhatfit}
\gsqb \ll 1 \; \mbox{ limit of }
\Cbp =   - \frac{\gfour}{8\pi} \bbp
- \frac{\gsq}{6\pi} \ln(\bbp \mD) \mD^2 \bps
  + \frac{\hat{q}}{4} \bps + \OO(\bpc) .
\end{equation}
Figure \ref{small_gsqb} plots the small
$\bbp$ data with two fits.  The long-dashed curves are based on
assuming that $\hat{q}$ is determined by the NLO expression presented
in \Eq{CNLO} -- that is, neglecting the (nonperturbative) $\gsix$
corrections.  The dashed curves represent a fit of the data, using all
data points with $\gsqb \leq 0.75$, to the functional form shown in
\Eq{qhatfit}, treating $\hat{q}$ as a free fitting coefficient and
neglecting $\OO(\bpc)$ effects (a
one-parameter fit).  The resulting value for $\hat{q}$
(which, note, is corrected by the $\ln(\bbp\mD)$ term), appears as
an added line of Table \ref{sim_res_tab}.  For the smallest three temperatures, 
this result is about a factor of 2 smaller than the NLO result, indicating that 
in these cases the next correction is not small. However, the deviation at the 
largest temperature shrinks to a factor of $1.3$. Recall that this was the only 
scenario in which $y \gg 1$ is actually fulfilled and convergence of the perturbative
series in $1/ \sqrt{y}$ is not hopeless. Even better agreement is expected at larger 
temperatures, where the first few orders in the perturbative expansion
should show convergent behavior.

\begin{figure}[phtb] 
\centering 
\includegraphics[width=0.8\textwidth,keepaspectratio]{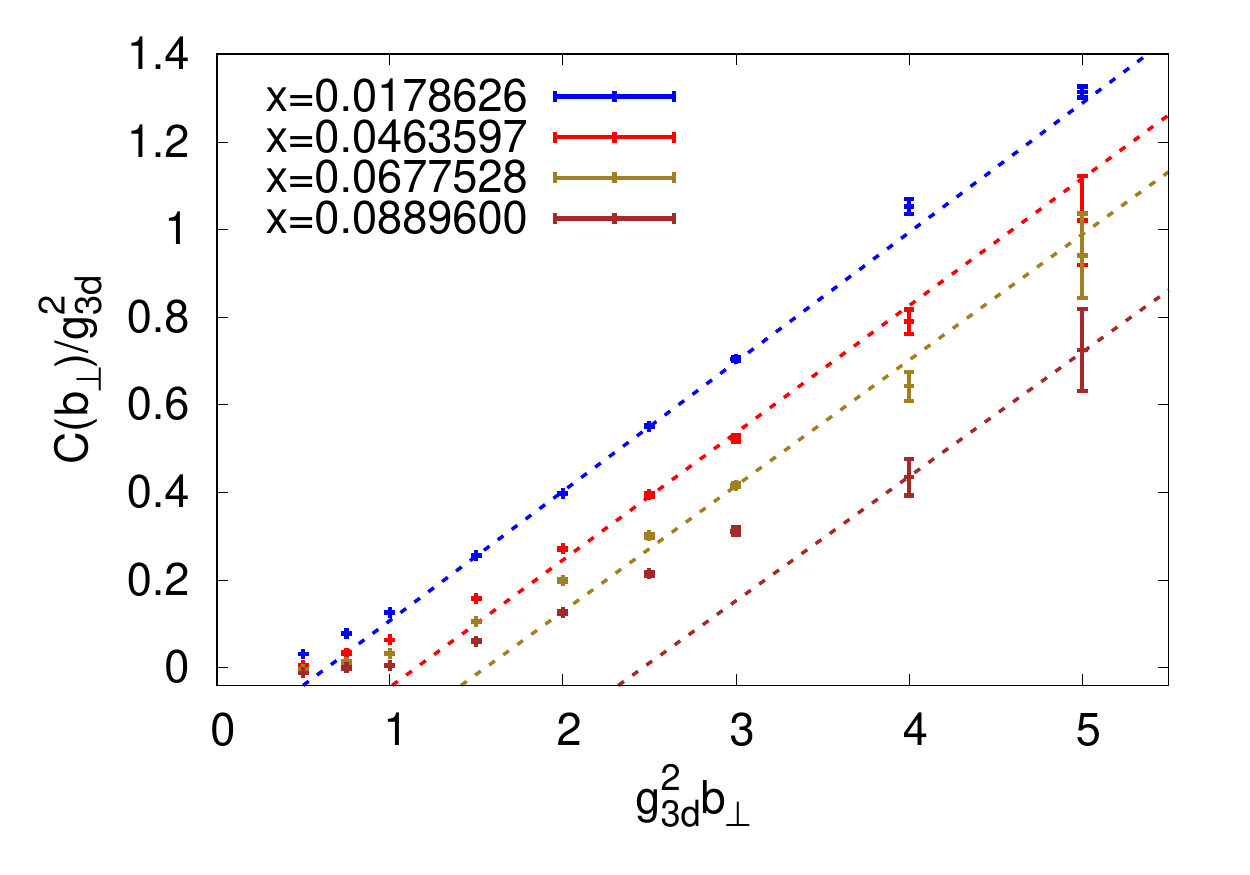}
\caption{Large $\bbp$ behavior of $\Cbp$, compared to straight-line
  asymptotics based on the string tension (see text).}
	\label{asy_cont_limits}
\end{figure}

Finally, we consider the large $\bbp$ asymptotics.  As discussed, one
expects $\Cbp$ at large $\bbp$ to rise linearly, with a coefficient
which equals the string tension. We list the string tension of EQCD 
in the last separated line of Tab~\ref{sim_res_tab}, determined by 
Laine and Schröder by perturbatively promoting the MQCD results of Teper
\cite{Teper:1998te} to full EQCD \cite{Laine:2005ai}. Just as in the small
$\gsqb$-region, a series expansion in $1/\sqrt{y}$ is involved, although 
this particular one seems to be a little bit more robust than the one for 
$\hat{q}$.%
\footnote{%
  It is perhaps not surprising that the perturbative series in
  $\sqrt{y}$ works better for the string tension than for $\hat{q}$.
  First, the string tension is determined solely by the gauge
  (magnetic) degrees of freedom, while $\hat{q}$ involves both
  electric and magnetic fields.  Second, there is already
  nonperturbative input in $\sigma$, from solving the gauge sector,
  while our information about $\hat{q}$ is only the short-distance
  perturbative contribution.}
The figure shows that the large-$\bbp$ asymptotics are indeed well described by the string
tension in EQCD.  The range over which this linear behavior holds is
larger for the high-temperature (small-$x$, large-$y$) case, where the
scalar decouples over a shorter distance.  For our
highest-temperature case we find nearly linear behavior already
close to $\gsqb \simeq 1$.  For the lowest temperature, on the other
hand, the scalar is very light and its effects persist to larger
distances; here it takes until approximately
$\gsqb \simeq 4$ before the asymptotic behavior sets in.
Due to the late set-in at the smallest temperature and the large statistical errors of the corresponding data points, 
we cannot really confirm the prediction of \Eq{EQCD_string_tension} here, even though the qualitative behavior, 
namely a linear asymptotic form, is nevertheless displayed by our data.
For the three higher temperatures, we find that the predicted value for the string tension 
fits our data very well.

\section{Conclusion and outlook}
\label{conclusion}

The collision kernel $\Cbp$ contains essential information about how a
thermal medium transverse-broadens, and therefore damps, high energy
particles.  Its perturbatively ill-behaved infrared contributions can
be computed within the effective theory EQCD by formulating the
collision kernel in terms of EQCD variables \cite{CaronHuot:2008ni}
and computing them on the lattice \cite{Panero:2013pla}.  We have
presented the first such lattice analysis of $\Cbp$ which is complete
in the sense that it uses all lattice-continuum improvements and makes
a complete extrapolation to the continuum limit.  Our approach
profited from the use of the multilevel algorithm for noise reduction
and the variable projection method for fitting to the long Wilson-loop
limit.  The use of fully improved lattice-continuum matching and
operator definitions accelerated the continuum limit, leading to high
precision continuum-extrapolated results.

Our results indicate a rather large downward correction in the value
of $\hat{q}$ relative to the NLO result at temperatures which are in 
the region of a few hundred $\mathrm{MeV}$ to several $\mathrm{GeV}$; 
indeed, we find a $\hat{q}$ value which is closer to the leading-order 
result.  This has implications for jet quenching and for transport coefficients, which
are also sensitive to $\hat{q}$.  However we want to emphasize here
that the reader should not dwell on the value of $\hat{q}$ itself,
which is a crude fit to a few small-$\bp$ points.  Rather, interested
practitioners should directly use $\Cbp$, which is the potential, in
the transverse plane, which is relevant for driving decoherence within
Zakharov's spacetime picture of jet quenching
\cite{Zakharov:1996fv,Zakharov:1997uu,Zakharov:1998sv,Baier:2000mf}.

Another interesting feature of our result is that the
value of the kernel $\Cbp$ is actually negative for sufficiently small
$\gsqb$ values.  This behavior is both expected and confusing.  In
Ref.~\cite{Ghiglieri:2013gia} it is shown that $\Cbp$ should have a
small-$\bbp$ expansion displaying a leading negative,
linear-in-$\bbp$ behavior.
A negative value of $\Cbp$ does not have a sensible probabilistic
interpretation; but this negative linear term can only dominate the
result for $\bbp \sim \gsq / \mD^2$, which is formally
$\OO(1/T)$.  This is exactly the short-distance regime where EQCD
breaks down as an effective description of thermal QCD.
Unfortunately, short distance, which corresponds to high transverse
momentum, is relevant for the highest-energy splitting processes,
which may be physically important in the medium-modification of high
energy jets.  Therefore our results need to be supplemented with a
matching calculation between the value of $\Cbp$ in EQCD and in full
QCD.  Most of the steps in this matching procedure have already been
taken \cite{Arnold:2008vd,CaronHuot:2008ni,Ghiglieri:2018ltw}.
A complete leading-order matching will allow our results to be
correctly incorporated into jet-quenching calculations.  A
higher-order analysis would shed
more light into the role of collinear effects, which are expected to
enter at the NNLO level but should be enhanced by double logarithms
\cite{Iancu:2015vea} and may involve more complex structures than the
Wilson loop considered here
\cite{Caron-Huot:2013fea,Caron-Huot:2015bja}.  We leave this, and an
application of our results to the computation of jet modification, for
future work.

\section*{Acknowledgments}
This work was supported by the Deutsche Forschungsgemeinschaft (DFG,
German Research Foundation) – project number 315477589 – CRC TRR
211. Calculations for this research were conducted on the Lichtenberg
high performance computer of the TU Darmstadt. We thank Kari
Rummukainen and Aleksi Kurkela for useful conversations, Harvey
B. Meyer, Hauke Sandmeyer and Olaf Kaczmarek for their advice on the
multilevel algorithm and Parikshit Junnarkar for drawing our attention
to the procedure of variable projection.

\appendix
\section{Simulation parameters and correlation matrices}
\label{app1}

For completeness, we provide a ``data dump'' of the details of our
simulated boxes, statistics, and correlation matrices.  Because the
correlation matrices are symmetric and nearly band-diagonal, we will
only list the entries 1 to 3 places to the right of the diagonal.

We also explain what at first sight is an odd choice of lattice volumes.
The volume should be chosen such that
$N_{x/y/z} \geq \beta$ in order to avoid finite volume effects
\cite{Hietanen:2008tv}. Furthermore, it has to fulfill $N_{x/y} > 2
\bbp^\mathrm{max}$ and $N_z > 2 L^\mathrm{max}$ to suppress an
(unphysical) correlation over the boundaries of the simulated
box. Since not all $\gsqa$ lattices produce sensible information at
all $\gsqb$ and only 3 lattices are required for a continuum limit,
the choice of volumes seems a little odd at first sight.

\begin{table}[htbp!] 
\centering {\footnotesize
\begin{tabular}{|c|c|c|c|m{2.4cm}|m{2.0cm}|c|}	
\hline
$\gsqa$ & $\xc$ & $\yc$ & $N_\mathrm{x} N_\mathrm{y}
N_\mathrm{z}$
& $\bp / a$ & $ L / a$ & statistics\\
\hline
$1/4$ & $0.08896$ & $0.452423$ & $52^2 \times 64$ & $4,6,8,10,12,$ $16,20,24$ & $4,8,12,16,$ $20,24,28$ & $7840$ \\
$1/6$ & $0.08896$ & $0.452423$ & $76^2 \times 96$ & $6,9,12,15,18,$ $24,30,36$ & $6,12,18,24,$ $30,36,42$ & $5900$ \\
$1/8$ & $0.08896$ & $0.452423$ & $100^2 \times 128$ & 
$4,6,8,12,16,20,$ $24,32,40,48$ & $8,16,24,32,$ $40,48,56$ & $4760$ \\
$1/12$ & $0.08896$ & $0.452423$ & $72^2 \times 192$ & $6,9,12,18,$ $24,30$ & $12,24,36,48,$ $60,72,84$ & $4000$ \\
$1/16$ & $0.08896$ & $0.452423$ & $96^2 \times 256$ & $4,8,12,$ $16,24$ & $16,32,48,64,$ $80,96,112$ & $5480$ \\
$1/24$ & $0.08896$ & $0.452423$ & $144^2 \times 384$ & $6,12$ & $24,48,72,96,$ $120,144,168$ & $240$ \\
$1/32$ & $0.08896$ & $0.452423$ & $192^2 \times 512$ & $4,8,16$ & $32,64,96,128,$ $160,192,224$ & $60$ \\
\hline
$1/4$ & $0.0677528$ & $0.586204$ & $52^2 \times 64$ & $4,6,8,10,12,$ $16,20,24$ & $4,8,12,16,$ $20,24,28$ & $8740$ \\
$1/6$ & $0.0677528$ & $0.586204$ & $76^2 \times 96$ & $6,9,12,15,18,$ $24,30,36$ & $6,12,18,24,$ $30,36,42$ & $5700$ \\
$1/8$ & $0.0677528$ & $0.586204$ & $100^2 \times 128$ & $4,6,8,12,16,20,$ $24,32,40,48$ & $8,16,24,32,$ $40,48,56$ & $2800$ \\
$1/12$ & $0.0677528$ & $0.586204$ & $72^2 \times 192$ & $6,9,12,18,$ $24,30$ & $12,24,36,48,$ $60,72,84$ & $4560$ \\
$1/16$ & $0.0677528$ & $0.586204$ & $96^2 \times 256$ & $4,8,12,$ $16,24$ & $16,32,48,64,$ $80,96,112$ & $5540$ \\
\hline
$1/4$ & $0.0463597$ & $0.823449$ & $52^2 \times 64$ & $4,6,8,10,12,$ $16,20,24$ & $4,8,12,16,$ $20,24,28$ & $8600$ \\
$1/6$ & $0.0463597$ & $0.823449$ & $76^2 \times 96$ & $6,9,12,15,18,$ $24,30,36$ & $6,12,18,24,$ $30,36,42$ & $4660$ \\
$1/8$ & $0.0463597$ & $0.823449$ & $100^2 \times 128$ & $4,6,8,12,16,20,$ $24,32,40,48$ & $8,16,24,32,$ $40,48,56$ & $3790$ \\
$1/12$ & $0.0463597$ & $0.823449$ & $72^2 \times 192$ & $6,9,12,18,$ $24,30$ & $12,24,36,48,$ $60,72,84$ & $4600$ \\
$1/16$ & $0.0463597$ & $0.823449$ & $96^2 \times 256$ & $4,8,12,$ $16,24$ & $16,32,48,64,$ $80,96,112$ & $5820$ \\
\hline
$1/4$ & $0.0178626$ & $1.64668$ & $52^2 \times 64$ & $4,6,8,10,12,$ $16,20,24$ & $4,8,12,16,$ $20,24,28$ & $7760$ \\
$1/6$ & $0.0178626$ & $1.64668$ & $76^2 \times 96$ & $6,9,12,15,18,$ $24,30,36$ & $6,12,18,24,$ $30,36,42$ & $6500$ \\
$1/8$ & $0.0178626$ & $1.64668$ & $100^2 \times 128$ & $4,6,8,12,16,20,$ $24,32,40,48$ & $8,16,24,32,$ $40,48,56$ & $4780$ \\
$1/12$ & $0.0178626$ & $1.64668$ & $72^2 \times 192$ & $6,9,12,18,$ $24,30$ & $12,24,36,48,$ $60,72,84$ & $2920$ \\
$1/16$ & $0.0178626$ & $1.64668$ & $96^2 \times 256$ & $4,8,12,$ $16,24$ & $16,32,48,64,$ $80,96,112$ & $4080$ \\
$1/24$ & $0.0178626$ & $1.64668$ & $144^2 \times 384$ & $6,12$ & $24,48,72,96,$ $120,144,168$ & $240$ \\
$1/32$ & $0.0178626$ & $1.64668$ & $192^2 \times 512$ & $4,8,16$ & $32,64,96,128,$ $160,192,224$ & $60$ \\
\hline
\end{tabular} }
\caption{Parameters for all EQCD multi-level simulations.}
\label{sim_params}
\end{table}

\begin{table}[htbp!] 
\centering {\footnotesize
\begin{tabular}{|c|c c c|}	
\hline
$\gsqb$ & nearest neighbor & next & next \\
\hline
$0.25$ & $0.29$ & $0.012$ & $0.0033$ \\
$0.5$ & $0.0095$ & $0.016$ & $0.0036$ \\
$0.75$ & $0.73$ & $0.17$ & $0.19$ \\
$1.0$ & $0.32$ & $0.31$ & $0.22$ \\
$1.5$ & $0.24$ & $0.47$ & $0.63$ \\
$2.0$ & $0.47$ & $0.13$ & $0.014$ \\
$2.5$ & $0.094$ & $0.060$ & $0.048$ \\
$3.0$ & $0.48$ & $0.28$ & $-0.18$ \\
$4.0$ & $0.38$ & $-0.049$ & - \\
$5.0$ & $-0.13$ & - & - \\
\hline
\end{tabular} }
\caption{Correlation matrix for $x=0.08896$ case.}
\label{corr_mat_x_0_08896}
\end{table}

\begin{table}[htbp!] 
\centering {\footnotesize
\begin{tabular}{|c|c c c|}	
\hline
$\gsqb$ & nearest neighbor & next & next \\
\hline
$0.5$ & $0.60$ & $0.65$ & $0.58$ \\
$0.75$ & $0.72$ & $0.91$ & $0.36$ \\
$1.0$ & $0.47$ & $0.46$ & $0.46$ \\
$1.5$ & $0.27$ & $0.37$ & $0.12$ \\
$2.0$ & $0.31$ & $0.13$ & $0.24$ \\
$2.5$ & $0.30$ & $0.03$ & $0.036$ \\
$3.0$ & $0.036$ & $0.09$ & $0.088$ \\
$4.0$ & $-0.029$ & $-0.011$ & - \\
$5.0$ & $-0.0041$ & - & - \\
\hline
\end{tabular} }
\caption{Correlation matrix for $x=0.0677528$ case.}
\label{corr_mat_x_0_0677528}
\end{table}

\begin{table}[htbp!] 
\centering {\footnotesize
\begin{tabular}{|c|c c c|}	
\hline
$\gsqb$ & nearest neighbor & next & next \\
\hline
$0.5$ & $043$ & $0.66$ & $0.26$ \\
$0.75$ & $0.64$ & $0.53$ & $0.55$ \\
$1.0$ & $0.32$ & $0.15$ & $0.26$ \\
$1.5$ & $0.61$ & $0.46$ & $0.13$ \\
$2.0$ & $0.54$ & $0.13$ & $0.0082$ \\
$2.5$ & $0.13$ & $-0.066$ & $0.039$ \\
$3.0$ & $0.03$ & $0.094$ & $-0.21$ \\
$4.0$ & $0.17$ & $-0.035$ & - \\
$5.0$ & $-0.071$ & - & - \\
\hline
\end{tabular} }
\caption{Correlation matrix for $x=0.0463597$ case.}
\label{corr_mat_x_0_0463597}
\end{table}

\begin{table}[htbp!] 
\centering {\footnotesize
\begin{tabular}{|c|c c c|}	
\hline
$\gsqb$ & nearest neighbor & next & next \\
\hline
$0.25$ & $0.17$ & $-0.26$ & $-0.53$ \\
$0.5$ & $0.027$ & $0.032$ & $0.023$ \\
$0.75$ & $0.82$ & $0.34$ & $0.65$ \\
$1.0$ & $0.39$ & $0.20$ & $0.015$ \\
$1.5$ & $0.66$ & $0.037$ & $0.04$ \\
$2.0$ & $0.13$ & $-0.0037$ & $0.096$ \\
$2.5$ & $-0.0017$ & $0.02$ & $0.0054$ \\
$3.0$ & $0.037$ & $0.0066$ & $0.0013$ \\
$4.0$ & $0.034$ & $-0.18$ & - \\
$5.0$ & $0.032$ & - & - \\
\hline
\end{tabular} }
\caption{Correlation matrix for $x=0.0178626$ case.}
\label{corr_mat_x_0_0178626}
\end{table}

\FloatBarrier
%\newpage
\bibliographystyle{unsrt}
\bibliography{references}

\end{document}